\documentclass[aps,prl,twocolumn,showpacs,amsmath,amssymb,preprintnumbers,nofootinbib,reprint,10pt]{revtex4-1}
\usepackage{graphicx}
\usepackage{bm}
\usepackage[pdftex]{hyperref}\hypersetup{colorlinks=true,linktoc=all,linkcolor=blue,breaklinks=true,citecolor=blue,urlcolor=blue}
\usepackage{natbib}
\usepackage{url}
\usepackage{color}

\begin{document}

\title{Observation of broadband entanglement in microwave radiation from a single time-varying boundary condition}
\author{B.H. Schneider}
\author{A. Bengtsson} 
\author{I.M. Svensson}
\author{T. Aref} 
\author{G. Johansson} 
\author{Jonas Bylander} 
\author{P. Delsing}
\affiliation{Department of Microtechnology and Nanoscience (MC2), Chalmers University of Technology, SE-412 96 Gothenburg, Sweden}

\date{\today}

\begin{abstract}
Entangled pairs of microwave photons are commonly produced in the narrow frequency band of a resonator, which represents a modified vacuum density of states. We use a broadband, semi-infinite transmission line terminated by a superconducting quantum interference device (SQUID). A weak pump signal modulates the SQUID inductance, resulting in a single time-varying boundary condition. We detect both quadratures of the microwave radiation emitted at two different frequencies separated by 0.7~GHz. We determine the type and purity of entanglement from the noise correlations and an in-situ noise and power calibration.
\end{abstract}

\maketitle
A time-varying boundary condition for the electromagnetic field can generate entangled photon pairs from the quantum vacuum. This fundamental property is called the dynamical Casimir effect (DCE) \cite{Moore1970}.
At low temperatures, the resulting output radiation due to the DCE exhibits two-mode squeezing \cite{Wilson2011}, which means that a portion of the noise is shared between two modes. Whenever the ratio of the shared noise to the non-shared noise in two detected modes is larger than a certain threshold \cite{Duan2000Mar}, the two modes are quantum entangled. Quantum back-action, e.g. due to a projective measurement of one mode, then affects both modes. 

%With the development in the field of quantum electro dynamics and quantum information processing \cite{Gu2017Nov}, 
Sources of entangled optical photons have been used in quantum secure key distribution \cite{Ekert1991Aug}, quantum repeaters \cite{Briegel1998Dec}, and quantum sensing applications \cite{Dixon2011}. 
At microwave frequencies, two-mode entanglement was proposed for entangling qubits \cite{Felicetti2014}, for continuous-variable quantum computing \cite{Andersen2015} and quantum enhanced detection at ambient conditions \cite{Barzanjeh2015Feb, Barzanjeh2018Sep}.
Sources of microwave entanglement such as parametric amplifiers \cite{Eichler2011, Zhong2013, Chang2017Aug}and modulated non-linear media \cite{Lahteenmaki2013} are comprised of a time-varying boundary condition or light velocity within a cavity; this arrangement enhances the radiation within the relatively narrow bandwidth of the cavity but suppresses it outside of this band.
In contrast, broadband entanglement sources are not as common \cite{Forgues2015Apr} but are useful for two reasons: i) they can be very bright and generate a large number of entangled photons and ii) their wide frequency content allows for shaping of the emitted radiation in time.
Good temporal control over the photon generation process is required in order to shape photon packages \cite{Rego2014, Silva2011}.
Protocols that reach unity efficiency in transmitting and absorbing photons rely on such temporal shaping \cite{Cirac1997, Jahne2007, Korotkov2011, Pechal2014}.

Broadband two-mode squeezing of microwave radiation was demonstrated by means of the DCE in a transmission line \cite{Wilson2011}.
However, imprecision in the determination of the output photon flux and a non-linearity due to strong pumping precluded the unequivocal demonstration of entanglement between photons.
Since then, quantitative bounds for entanglement have been developed \cite{Johansson2013}, taking thermal photons into account.
%into account thermal photons.
%More recent theoretical advances are currently paving towards a non-perturbative treatment of the dynamical Casimir effect taking possible mechanical elements into account \cite{Macrn2018Feb}.

\begin{figure}[!t!]
  \centering
  \includegraphics{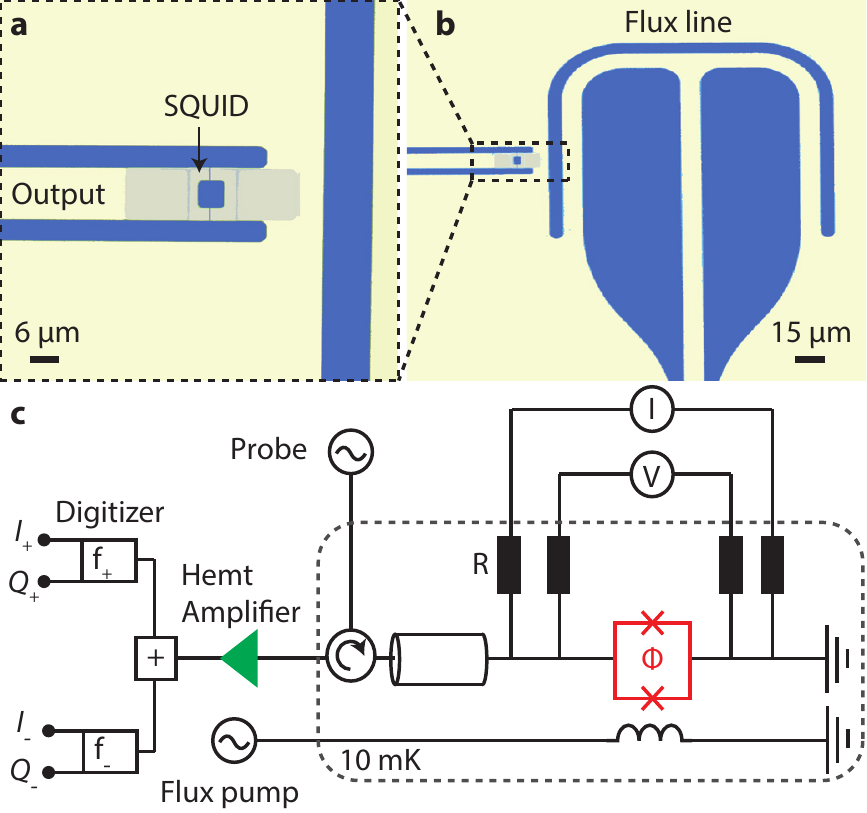}
  \caption{False-color photograph and simplified circuit schematic of the measured sample. (a,b), The DC-SQUID is made of aluminum (grey) and connects a coplanar waveguide transmission line to ground.
  The transmission line and the ground plane are made of niobium (yellow) on top of a sapphire substrate (blue).
  (b), On-chip magnetic flux line located next to the SQUID.
  (c), Simplified circuit diagram showing cables carrying alternating and direct currents to the device.}
  \label{fig:1}
\end{figure}

In this paper we demonstrate broadband entanglement of microwave photon pairs generated by the DCE in a superconducting circuit.
The circuit consists of a semi-infinite transmission line, terminated by two parallel Josephson junctions connected to ground (a direct-current superconducting quantum interference device, DC-SQUID); see Fig.~1(a).
This SQUID's Josephson inductance represents a variable boundary condition or "movable mirror" for the electromagnetic field \cite{Johansson2009Sep}. We rapidly modulate this boundary condition, at a microwave frequency ($f_p$ = 8.9~GHz), by means of pumping the magnetic flux threading the SQUID loop, thereby producing DCE radiation.
We detect the in-phase and quadrature voltages of the output field of the transmission line at different pump amplitudes, i.e. at different displacement speeds of the electromagnetic boundary condition. We then compute the covariance matrix of the voltage fluctuations at two different frequencies $f_+$ and $f_-$ (where $f_+ + f_- = f_p$ and $f_+ - f_- = 0.7$~GHz), and further determine the log-negativity and the purity of entanglement; we calculate the amount of two-mode squeezing below the vacuum level, and from this we determine the type of entanglement.
This quantification of entanglement relies on our careful calibration of the system gain and noise level and the flux-pump amplitude \cite{appendix}.
Measurements were done in a dilution refrigerator at a temperature of 10~mK.
The device under test consists of an aluminum DC-SQUID with a loop area of 6x8~$\mu \text{m}^2$. It is directly connected to a 0.6~mm long on-chip superconducting niobium transmission line. The ground plane and flux-pump antenna are also made in the same niobium layer. A semiconductor HEMT(high electron mobility transistor) amplifier with 39~dB gain amplifies the signals in the range of 4-8~GHz.
After additional amplification and filtering, two digitizers detect the heterodyne down-converted signals at two frequencies, yielding the quadrature voltages $I$ and $Q$ at frequencies $f_+ = 4.8$~GHz and $f_- = 4.1$~GHz.
A probe signal can be launched via the circulator and is used to characterize the change in phase and magnitude of the reflected signal.
Furthermore, four low-pass-filtered wires are connected across the SQUID to enable the DC characterization.
The low-pass filtering in these lines consists of high resistance-capacitance and copper-powder filters with a total cut-off frequency of 30~Hz.
An external magnetic coil is used to set the static flux ($\Phi_{DC}$) of the SQUID.
We modulate the boundary condition by sending an AC signal ($\Phi_{AC}$) to an on-chip flux pump line (Fig.~\ref{fig:1}b).
The flux-pumping frequency can be chosen arbitrarily; here we present data for $f_p = f_+ + f_- = 8.9$~GHz.

\begin{figure*}[!t!]
  \centering
  \includegraphics{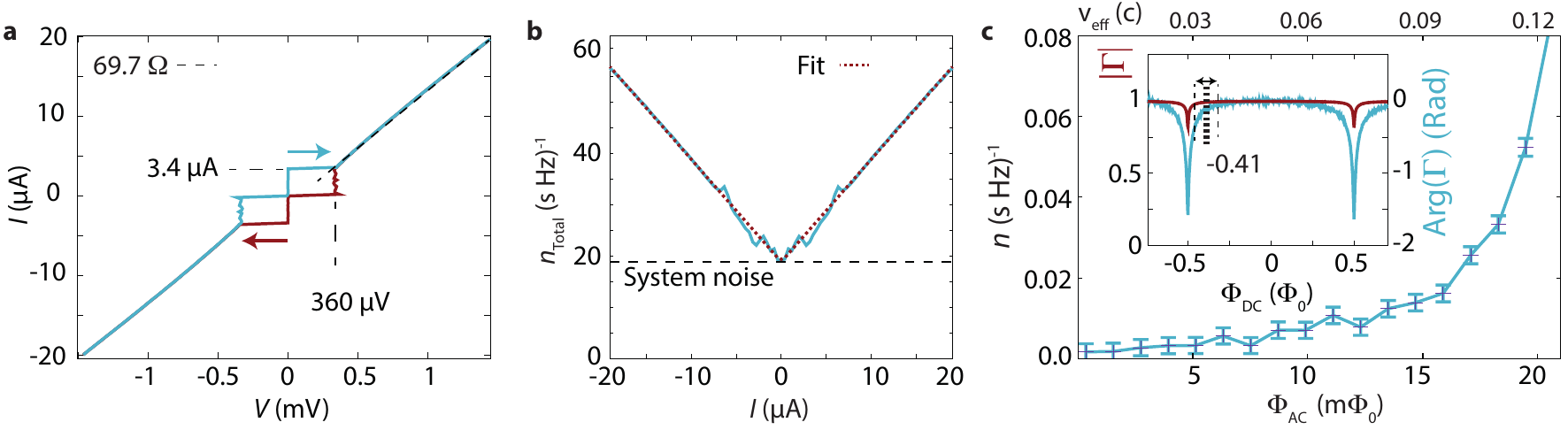}
  \caption{SQUID characterization and circuit calibration. (a), Current-voltage characteristic of the SQUID
    (up-sweep blue, down-sweep red). 
    The high-bias resistance is $69.7~\Omega$ (dashed line).
    (b), Shot noise photon spectral density at 4.1~GHz versus applied current.
    (c), Average photon spectral density generated as a function of flux-pump amplitude. 
    The top axis indicates the effective speed of the electromagnetic boundary condition relative to the speed of light.
    The inset shows the microwave reflected magnitude (red) and relative phase (blue) of a probe signal at 4.1~GHz.
    A vertical dashed line at $-0.41\Phi_{DC}$ indicates the static flux position used and the effective flux modulation range for 20~m$\Phi_0$ is also indicated.}
  \label{fig:2}
\end{figure*}

We first characterize the device by measuring the current-voltage characteristic of the SQUID (Fig.~\ref{fig:2}a) and find a critical current of $I_c = 3.4$~$\mu$A and a superconducting energy gap voltage of 360~$\mu$eV. 
From the forward (blue) and backwards (red) current sweeps, 
we observe a hysteresis, indicating that the SQUID is underdamped, with $\beta_C \approx 4I_c/ (\pi I_r) \approx 10^4$, where $I_r$ is the superconducting retrapping current.

To obtain the necessary resolution in the measurement of the voltage output from the transmission line, we use the SQUID itself to calibrate the noise and gain of the amplifier chain (Fig.~\ref{fig:2}b).
By applying a current through the SQUID, shot noise is generated \cite{Koch1980Dec}, which can be used to calibrate the system \cite{Spietz2003, Spietz2006}. At the same time, the voltage drop across the SQUID is measured, determining its resistance.
The resistance of the SQUID for a voltage above the gap, $V_g = \frac{2\Delta}{e}$, is $R = 69.7~\Omega$ (Fig.~\ref{fig:2}a).
The difference compared to the impedance of the transmission line, $Z_0 = 50~\Omega$, is taken into account using the following equations:
\begin{eqnarray}
  \! E_1 &=& \frac{V_{s}^2 + V_{z}^2}{V_{T}^2} \label{eq:E1}, \nonumber\\ 
  \! E_2 &=& \frac{V_{s}^2 - V_{z}^2}{V_{T}^2} \label{eq:E2}, \nonumber\\ 
 \! \! S_{p} &=& G Bw  \! \left[\frac{V_{T}^2}{2 Z_0} \left(\frac{E_1}{ \tanh(E_1)}\! + \!\frac{E_2}{ \tanh(E_2)}\right) + k_{\text{B}} T_n \right] \label{eq:shotnoise} \!\! ,
\end{eqnarray}
where $k_{\text{B}}$ is Boltzmann's constant, $T$ is the device temperature, $T_n$ is the system noise temperature referred to the device, and $G$ and $Bw$ are the gain and the detection bandwidth, respectively.
$V_{s}^2 = 2 e |I| R^2 \cdot Z_0^2/(Z_0 + R)^2  $, $V_{T}^2 =  4 k_{\text{B}} T \cdot Z_0^2/(Z_0 + R) $ and $V_{z}^2 = Z_0 \cdot \frac{1}{2} h f$ are spectral densities, which relate to the shot noise, Johnson noise, and zero-point fluctuations at frequency $f$, respectively.  $h$ is Planck's constant.
$S_{p}$ is the measured shot noise power spectral density. 

In Fig.~\ref{fig:2}b, we show the spectral density of the shot noise and the corresponding fit as a function of DC current through the SQUID, with a static magnetic flux of $\Phi_{DC} = -0.41~\Phi_{0}$, which is used throughout the paper. Here $\Phi_{0} = h/2e$ is the magnetic flux quantum. 
This fit accurately determines the system noise temperature and gain.
The system noise corresponds to a temperature of $3.71\pm0.04$~K at 4.1~GHz and $2.95 \pm 0.02$~K at 4.8~GHz, which matches the noise of the HEMT amplifier which is $2.3$~K at 4.1~GHz and 2~K~at 4.8~GHz, connected via two circulators and filters.
Since the HEMT amplifier dominates the noise, we can find the corresponding photon losses between device and amplifier.
Here we find a photon loss of  $10\log(2.2/3.7) \approx -2.3~$dB at 4.1~GHz and $10\log(2/2.95) \approx -1.7$~dB at 4.8~GHz.

To generate DCE photons, we apply a sinusoidal signal to the flux line at $f_p = f_- + f_+ = 8.9$~GHz, while recording the signal using two digitizers at $f_- = 4.1$~GHz and $f_+ = 4.8$~GHz, i.e., placed symmetrically around $f_p/2$ (Fig.~\ref{fig:1}c). DCE photons are generated in pairs symmetrically around half the pump frequency, thus by using a photon spectral density we can compare the photon rates.
The effective speed of "mirror" displacement is given by the phase response (inset Fig.~\ref{fig:2}c), the flux amplitude and $f_p$.
For small amplitudes, the phase depends linearly on the flux, such that the boundary condition can be mapped to a sinusoidally moving mirror.
With a flux pump amplitude $\Phi_{AC}$ exceeding 15~m$\Phi_0$, the change in phase becomes larger, which results in a larger photon spectral density; however, the motion also becomes non-linear \cite{appendix}.
A power calibration of the flux pump amplitude $\Phi_{AC}$ is shown in \cite{appendix}.

We experimentally track changes in the output radiation such as photons generated by the DCE. 
We do this by switching the pump on and off and tracking the difference.
From the previous calibration (Fig.~\ref{fig:2}a, b), 
we obtained a photon spectral density of $0.5 \pm 0.0035~(\text{s~Hz})^{-1}$ corresponding to the vacuum fluctuations, when the pump is switched off, and the system noise is subtracted.
The background noise in the system is determined by subtracting the amplifier noise and the zero-point fluctuations from the total input noise.
Any remaining noise signal would be due to thermal photons. This is smaller than what we could resolve, confirming a photon temperature of less than 40~mK.
However, the uncertainty in the background noise $\pm 0.0035$ (s Hz)$^{-1}$ is not small enough to resolve temperature below 40~mK, which corresponds to $n_{th} = 1/(\exp(\text{h}f/(k_{\text{B}}T)) -1) = 0.0031$ at 4.8~GHz.
As we increase the flux pump amplitude, we measure an increase in photon spectral density.
Figure \ref{fig:2}c, shows the generated photon spectral density versus flux pump amplitude.

We use two methods to probe and characterize entanglement between produced photon pairs: first, by calculating the log-negativity, and second, by comparing the quadrature noise to the vacuum.
Both methods are commonly used to probe entanglement and non-separability \cite{Simon2000, Adesso2007}.

\begin{figure*}[!ht!]
  \centering
  \includegraphics[width=\linewidth]{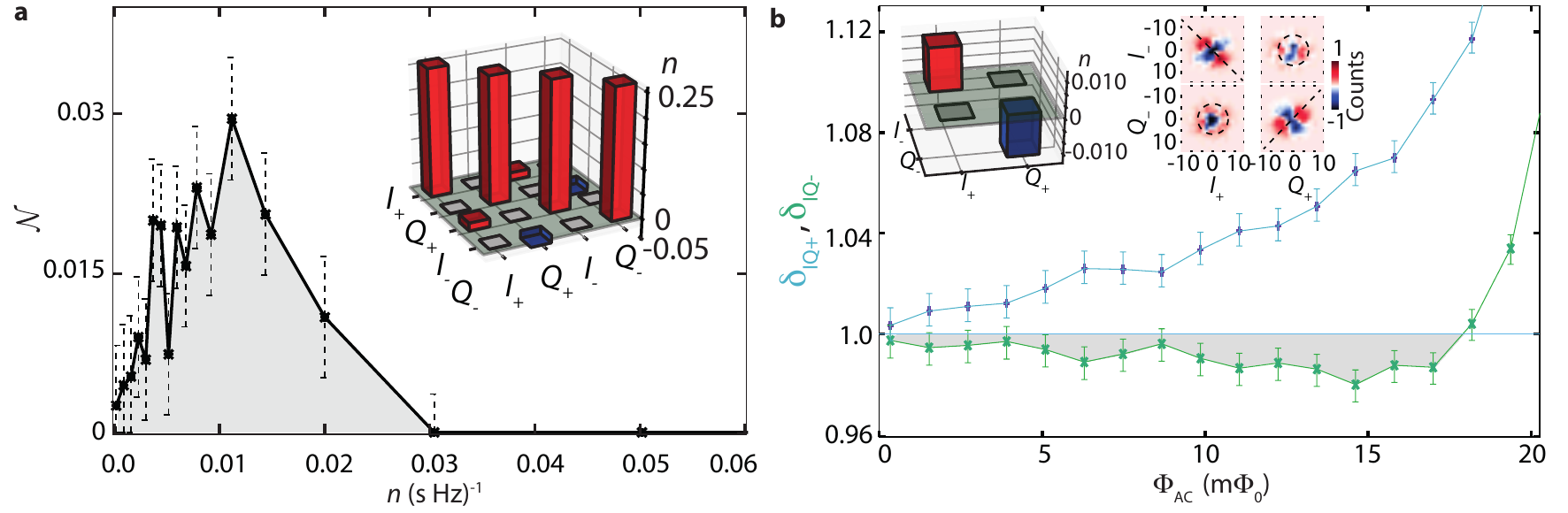}
\caption{Two measures of entanglement.
  (a), Logarithmic negativity versus photon spectral density $n$.
  $\mathcal{N} > 0$ (shaded grey) indicates entanglement. 
  The inset shows a covariance matrix, taken at a photon spectral density of $n = 0.01$, which is used to calculate the log-negativity.
(b), Combined quadrature fluctuations in the $I$ and $Q$ quadratures as a function of flux pump amplitude $\Phi_{AC}$.
At low amplitudes, we observe squeezing below the vacuum in both quadratures (shaded grey) when $\delta_{IQ-} < 1$, fulfilling the inseparability criterion \cite{Duan2000Mar}.
The two insets show the off-diagonal elements of the covariance matrix with the corresponding quadrature histograms. 
Each histogram is calculated from the difference between flux pump on and off.
}
\label{fig:3}
\end{figure*}

As we generate photons using the DCE, shown in Fig.~\ref{fig:2}c, 
we record the voltage quadratures $I_+$, $Q_+$, $I_-$ and $Q_-$ corresponding to the frequencies $f_+$ and $f_-$.
From the quadrature correlations, we can construct the covariance matrix (inset in Fig.~\ref{fig:3}a).
Error values are estimated for all elements of the covariance matrix as one standard deviation.
Once the covariance matrix is established, we calculate the logarithmic negativity \cite{Johansson2013}:
\begin{eqnarray}
  \mathcal{N} &=& \max{[0, - \log_2(2\nu_-)]} ,\label{eqn:logneg} \\
  \nu_- &=& [\zeta/2 - (\zeta^{2}- 4 \det V)^{1/2}/2]^{1/2} , \nonumber\\
  \zeta &=& \det~A + \det~B - 2\det~C , \nonumber \\
  V &=&  \frac{1}{2} \begin{pmatrix} A & C \\ C^{T} & B \end{pmatrix} ,\nonumber
\end{eqnarray}
where $V$ is the $4\times 4$ covariance matrix with the $2\times 2$ sub-matrices $A$, $B$, and $C$.
The logarithmic negativity is positive for a photon spectral density of $0.03~(\text{s~Hz})^{-1}$ or lower, as can be seen in Fig.~\ref{fig:3}a.

The logarithmic negativity is lower than the theoretical value \cite{Johansson2013} ($\mathcal{N} \approx 2\sqrt{n} = 0.2$).
Similarly to Ref. \cite{Ku2015Apr}, we include measurement noises and losses in the presented results.
Photon losses in the system and a small non-linearity \cite{appendix} in the SQUID result in lower cross-correlation values.
By taking the previously estimated photon losses into account we obtain $\mathcal{N} \approx 0.1$ at the device, which is still approximately a factor two lower than the theoretically expected value. The remaining factor of two can be explained by the presence of a non-linearity in the response of the SQUID inductance to a magnetic flux (Fig.~\ref{fig:2}c).
This non-linearity results in an effective pumping at higher harmonics such as $2 f_p = 17.8$~GHz, producing unentangled photons at $f_+$ and $f_-$.
For larger photon spectral densities and flux pumping, $\Phi_{AC} > 15$~m$\Phi_0$, $\mathcal{N}$ decreases due to this non-linearity (see simulation results in \cite{appendix}).

The right inset in Fig.~\ref{fig:3}b shows four histograms of measured $I$ and $Q$ quadratures, taken at a flux pump amplitude $\Phi_{AC}=13$~m$\Phi_0$.
The histograms show the difference between flux pump on and off.
The top left $I_-I_+$ histogram and bottom right $Q_-Q_+$ histogram show squeezing along the dashed diagonals that are orthogonal to each other: photons are amplified along the diagonal dashed line and are squeezed orthogonally to it.

From the quadrature correlations, we calculate the combined quadrature fluctuations $\delta_{IQ+} = \langle (I_+ + I_-)^2\rangle + \langle (Q_+ - Q_-)^2\rangle$ and $\delta_{IQ-} = \langle (I_+ - I_-)^2 \rangle + \langle (Q_+ + Q_-)^2\rangle$ as a function of flux pump amplitude (Fig.~\ref{fig:3}b), where  the later fulfils the inseparability criterion for continuous variable systems by Duan \cite{Duan2000Mar, Treps2005} for values below 1.
We observed $-0.09~\pm~0.02$~dB squeezing below the vacuum in $\langle(I_+-I_-)^2\rangle$ and $\langle(Q_++Q_-)^2\rangle$. We also observed an amplification of 0.25~$\pm$~0.02~dB in $\langle(I_++I_-)^2\rangle$ and $\langle(Q_+-Q_-)^2\rangle$ at a flux pump strength of $\Phi_{AC} = 15$~m$\Phi_{0}$.
For low flux pump powers in the more linear regime, both methods indicate entanglement. The two modes for which we find entanglement are streams of photons from the DCE, we conclude that these photon pairs are entangled.

To compare the entanglement generation, we calculate the entangled bits with the entropy of formation \cite{Giedke2003Sep} for a given logarithmic negativity of 0.03, which is $E_F=(1.6 \pm 0.3) \cdot 10^{-3}$ at the amplifier input \cite{appendix}.
This corresponds to an entanglement rate of $\sim 5.2$~Mebit/s, in turn corresponding to a distribution rate of entangled Bell pairs \cite{Flurin2012Oct}.
These numbers are substantially larger at the device.
There are two reasons for this: losses between the device and the amplifier and the limited bandwidth of the amplifier. 
Taking losses into account and including the full bandwidth between DC and the pump frequency, we estimate $E_F = 13 \cdot 10^{-3}$ available at the device, corresponding to an entanglement rate of $\sim 90$ Mebit/s. The entanglement rate at the device is high (order of magnitude higher) in comparison to other entanglement sources\cite{appendix}.

We demonstrated that photon pairs generated by the DCE without a cavity are entangled.
To our knowledge entanglement of a single time-varying boundary condition without the presence of a cavity has previously not been observed.

\paragraph*{Acknowledgements:}
We thank C.M. Wilson, W. Wieczorek, V. Shumeiko, and N. Treps for useful discussions on methods and entanglement.
We gratefully acknowledge financial support from the European Research Council, the European project PROMISCE, 
the Swedish Research Council, and the Wallenberg Foundation. J.B. acknowledges partial support by the EU under REA Grant Agreement No. CIG-618353.

%

%\end{document}
% SUPPLEMENTARY HERE: 

\onecolumngrid
\clearpage
\begin{center}
\section*{\textbf{Supplementary}}
\end{center}
%\end{onecolumngrid}
\twocolumngrid

% Reset counter and equation numbering ...
\setcounter{equation}{0}
\setcounter{figure}{0}
\setcounter{table}{0}
\setcounter{page}{1}
\makeatletter
\renewcommand{\theequation}{SE\arabic{equation}}
\renewcommand{\thefigure}{S\arabic{figure}}
\renewcommand{\thetable}{S\arabic{table}}
\renewcommand{\bibnumfmt}[1]{[SC#1]}
\renewcommand{\citenumfont}[1]{SC#1}

\begin{figure}[ht]
  \centering
  \includegraphics{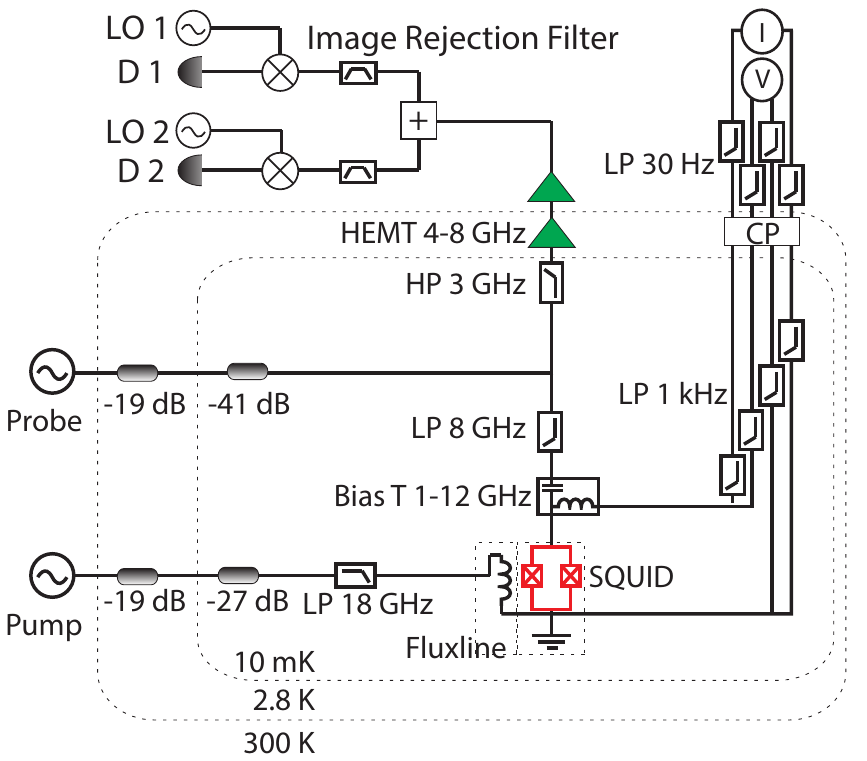}
  \caption{Circuit setup.
  Each DC line is filtered with a 30~Hz resistive low pass filters at room temperature, a copper powder filter at 2.8~K, and 1~kHz resistive low pass filters at 10~mK. The heavy filtering ensures electron thermalisation of the sample at base temperature.
  A wide band bias-T separates direct (DC) and alternating currents (AC), such that an AC signal propagating from the SQUID gets amplified by the HEMT amplifier.}
  \label{fig:circuit}
\end{figure}

\textbf{S1: Circuit diagram and pre calibration}\\
The detailed circuit diagram in Fig.~\ref{fig:circuit} shows the measurement setup used during the experiment.
Measurements were done in a dilution refrigerator at a temperature of 10 mK. The device under test consists of an aluminum DC-SQUID with a loop area of 6x8 $\mu$m$^2$. It is directly connected to a 0.6 mm long on-chip superconducting niobium transmission line. The ground plane and flux-pump antenna are also made in the same niobium layer. A semiconductor HEMT(high electron mobility transistor) amplifier with 39 dB gain amplifies the signals in the range of 4-8 GHz.
The 4 wire direct current (DC) measurements were done using 4 heavily filtered lines to ensure sample thermalisation at 10~mK.
To confirm sample temperature a slow and fine current sweep, revealing non-linear structures below the critical current, are fitted using the differential resistance of the device.
In this scenario we can confirm that the SQUID is thermalised with the fridge and thus is in the ground state ($kT < hf$).

During the measurements to allow in-situ calibration we execute short and fast current sweeps. 
For this we only fit the slope above the critical current, and by using and averaged, static resistance ($69.7~\Omega$). This this means, that we can run this short calibration interleaved with the measurements.\\

\textbf{S2: Estimation of errors}\\
By fitting the shot noise at each desired detector frequency, we get the individual amplification gains and system noise temperatures.
The ability to track drifts and capture changes to the system as they happen is used to achieve an upper bound on the noise present during the measurements.

The fitted results for two frequencies used $f_1$ = 4.8~GHz and $f_2$ = 4.1~GHz to the corresponding flux pump frequency $f_p = 8.9$~GHz (which is switched off during calibration), before and after the measurement are:
\begin{eqnarray}
  Gs_1 &=& 1.3051 \cdot 10^9 \pm 3.4 \cdot 10^6, \\
  Gs_2 &=& 1.4906 \cdot 10^9 \pm 3.6 \cdot 10^6, \\
  Ge_1 &=& 1.2929 \cdot 10^9 \pm 4.3 \cdot 10^6, \\
  Ge_2 &=& 1.4817 \cdot 10^9 \pm 5.6 \cdot 10^6,  
  \label{eqn:gains}
\end{eqnarray}
where $Gs_1$ and $Ge_1$ correspond to the gain at frequency $f_1$ at the start and at the end of the measurement
and $Gs_2$ and $Ge_2$ to the gains at frequency $f_2$.
This corresponds for a gain drift of 0.58~dB and 0.59~dB for $f_1$ and $f_2$ respectively over a time period of 8~hours.

The corresponding average values are:
\begin{eqnarray}
  (Gs_1 + Gs_2)/2 = G_{m1} &=& 1.299 \cdot 10^9, \\ %(1.30513199e+09 + 1.29291916e+09)/2.0 \\
  (Ge_1 + Ge_2)/2 = G_{m2} &=& 1.486 \cdot 10^9, \\ %(1.49057109e+09 + 1.48166961e+09)/2.0\\
  \Delta G_{m1} &=& 3.8 \cdot 10^6, \\ %(4263590+3405414)/2.0\\
  \Delta G_{m2} &=& 4.6 \cdot 10^6.    %(5601722+3612884)/2.0
\end{eqnarray}

We obtain the number of photons by dividing the power detected by the digitiser by a factor.
That depends on the respective frequency and gain, i.e. $B G_{m1} h f_1$ and $B G_{m2} h f_2$.
The photon numbers at the corresponding frequencies are:
\begin{equation}
  n_1 = \frac{ P_{n1,on} - P_{n1,off} }{B_w G_{m1} h f_1},\\
  \label{eqn:PhotNum1}
\end{equation}

\begin{equation}
  n_2 = \frac{ P_{n2,on} - P_{n2,off} }{B_w G_{m2} h f_2},
  \label{eqn:PhotNum2}
\end{equation}
where $P_{n1,on}$ and $P_{n1,off}$ are the detected powers at room temperature at frequency 1 (4.1~GHz) with the flux pump on or off, respectively.
Similarly, $P_{n2,on}$ and $P_{n2,off}$ are the powers at frequency 2 (4.8~GHz).
The signal of interest is the power difference between flux pump on and off ($P_{n,DCE}$).

To account for the error in this signal we can consider three aspects:
First, the uncertainty in the gain from the fit.
Second, the amount by which the gain drifted between two measurements.
Third, the overall noise present in $P_{n,off}$ with the same amount of averaging.
The error, due to uncertainty in the gain is:
\begin{equation}
  \Delta n = n \frac{\Delta G}{G_{m}},
  \label{eq:fituncertainty}
\end{equation}
where $\Delta n$ is the resulting uncertainty in photon numbers as a function of total photon numbers $n$ and the fraction of gain uncertainty $\frac{\Delta G}{G}$.
We find the amount of gain drift taking the difference between the start and end gains:
\begin{eqnarray}
  \Delta G_{Drift1} = Gs_1 - Ge_1 = 12.1 \cdot 10^6, \\  % 12.212830 \cdot 10^6 \\
  \Delta G_{Drift2} = Gs_2 - Ge_2 = 8.9 \cdot 10^6,   % 8.901480 \cdot 10^6
  \label{eq:Gain-Drift}
\end{eqnarray}
where $\Delta G_{Drift1}$ and $\Delta G_{Drift2}$ are our uncertainties in the gain due to drift. 
Together with the fit uncertainty we get:
\begin{eqnarray}
  \Delta G_1 = \sqrt{\Delta G_{Drift1}^2 + \Delta G_{m1}^2} = 12.8 \cdot 10^6, \\ %12800649
  \Delta G_2 = \sqrt{\Delta G_{Drift2}^2 + \Delta G_{m2}^2} = 10.0 \cdot 10^6. % 10023153
  \label{eq:gain_uncertainty}
\end{eqnarray}
The resulting uncertainty in the gain is given by:
\begin{eqnarray}
  \Delta G_1 / G_{m1} = 0.01, \\
  \Delta G_2 / G_{m2} = 0.007,
  \label{eq:Gu_percent}
\end{eqnarray}
which gives an overall gain accuracy within 1\%.

However, noise between two on-off cycles in succession might be more dominant than the uncertainty in the gain.
To investigate this, we calculate the variance for ($P_{n1,off}$ and $P_{n2,off}$) under same conditions as the measurements.
\begin{eqnarray}
  \Delta P_{n1} = \sqrt{var(P_{n1,off})} = 0.0025,\\
  \Delta P_{n2} = \sqrt{var(P_{n2,off})} = 0.0021,
  \label{eq:PnNoise}
\end{eqnarray}
which gives us an additional uncertainty of 0.0025 per photon in the OFF signal.
Assuming we have the same uncertainty when the flux pump is on and adding this, we get an uncertainty of $\sqrt{2}(0.0025) = 0.0035$ for n1 and $\sqrt{2}(0.0021) = 0.0029$ photons in n2.
Typically the number of photons in the differential signal in the region of interest is around 0.05 photons. This results on average in an statistical error of 6\% in the photon number resolution. Given time and equipment this statistical error could still be improved.\\

\textbf{S3: Flux pump power calibration}\\
\begin{figure}[ht]
\textbf{S3: Flux pump power calibration}
\centering
\includegraphics[width=\columnwidth]{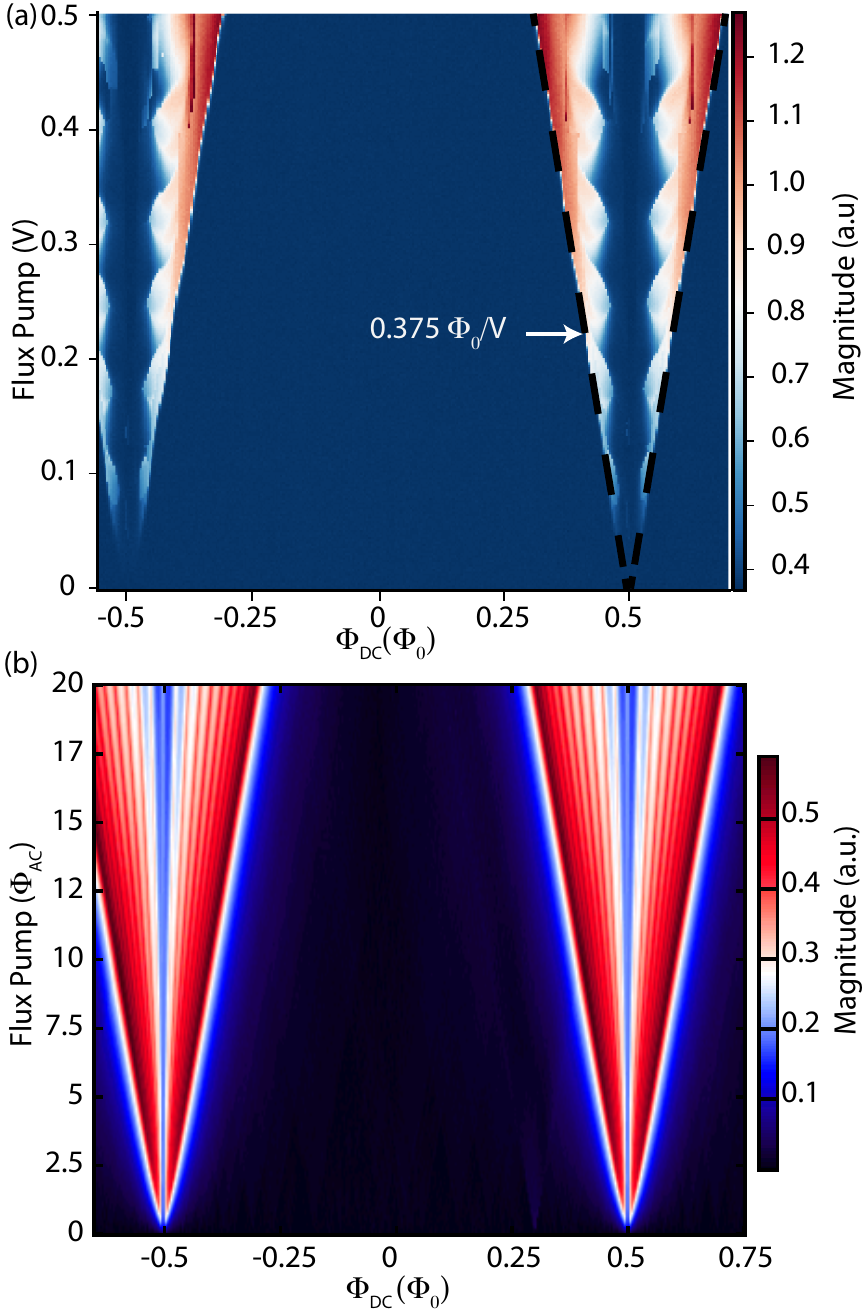}
\caption{Flux pump power calibration.
  (a) The color scale is the raw power detected by the digitiser at 4.1~GHz against DC and raw flux pump voltage.
  The black dashed line follows an onset of increased power.
  This slope of this onset is ($\Phi_{AC} = 0.375~\frac{\Phi_0}{V} \cdot V_{pump}$). This slope is in proximity of one half flux quanta, since the SQUID phase response is steepest, resulting in increased photon numbers.
(b) Simplified model of the expected signal, we can use this to estimate the non-linear response of the squid to the first order.}
\label{fig:pumpcalibration}
\end{figure}

Here we estimate the effective magnetic field through the SQUID as a function of flux pump strength. The signal generator is swept from 0 to 250~mW at room temperature. The magnetic field induced in the SQUID scales with the square root of the pump power. Therefore we sweep the output voltage of the signal generator.
The total magnetic flux through the SQUID is given by $\Phi_{AC}  + \Phi_{DC} $, where $\Phi_{AC}$ is the magnetic flux induced by the flux pump (which is proportional to the output voltage). The flux pump amplitude ($\Phi_{AC}$) acting on the SQUID, is estimated by fitting the onset for which the photon spectral density visibly increases as a function of flux pump amplitude and dc flux offset (Fig. \ref{fig:pumpcalibration}). This sudden increase in photon spectral density happens because the relative change in the SQUIDs inductance is largest at $0.5~\Phi_0$. 
Close to this point is where the mirror moves fastest.
The AC flux pump $\Phi_{AC}$ eventually reaches this point with increased pump amplitude, i.e. $\Phi_{AC}+\Phi_{DC} = 0.5 ~\Phi_0$ and this is where the photon generation increases drastically.
This sudden increase in photon numbers is then used to calibrate the flux pump strength.\\

%\newpage
\textbf{S4: Gaussianity and non-linearity of the system}\\
\begin{figure}[ht]
\centering
\includegraphics[width=\columnwidth]{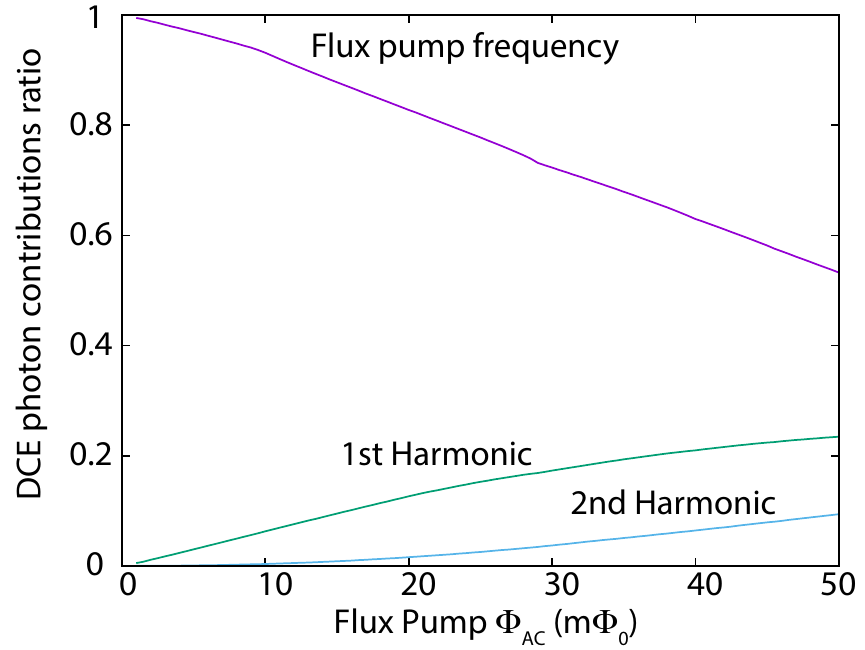}
\caption{DCE photon purity vs pump power. Using the model we can calculate the ratio of photons generated by higher harmonics. The effect of the non-linearity is that the measured squeezing of the device is reduced by contributions of higher harmonics.}
\label{fig:S4_NLratio}
\end{figure}

With the measured SQUID response (Fig. 2 (b) and 2 (c) of the main paper), we obtained enough information to model the system to the first order see fig. \ref{fig:pumpcalibration} (b). 
With only the first two orders of non-linear-response from a SQUID
we expect a purely Gaussian system.
However, photons generated by the second order non-linearity (from effective higher pump tones) do pollute the measured covariance matrix, effectively lowering the measured entanglement.
We here estimate the potential influence from the non-linear-response of SQUID, 
we quantify this by the ratio of photons generated from a linear response to a non-linear response shown in figure \ref{fig:S4_NLratio}.
(The python code including additional explanations is accessible here: \url{https://github.com/benschneider/Dynamical-Casimir-Effect-Sim}, whereas figure visualisation is done using Spyview\cite{spyview2017}.)

To obtain a non-Gaussian mode one needs to couple different modes, meaning that generated entangled DCE photons need to be involved in the generation of additional photons.
This can happen by reflections for example in a cavity and by higher order non-linear terms (above the second order) of the SQUID. 
The even non-linear terms in a SQUID scale with $Z_0/R_k \cdot \ln\left( 1+ Z_J/Z0\right) \sim 0.0017$, where $Z_0 = 50 \Omega$ is the impedance of the transmission line, $R_k=h/e^2\sim25 k\Omega$ is the quantum resistance and $Z_J=\sqrt{L_J/C_J}\sim70 \Omega$ is the impedance of the Josephson junctions.
Given this, and the absence of a cavity, it will be virtually impossible to obtain any non-Gaussian state, given a classical sinusoidal flux pump.

\textbf{S5: Entangled bits}\\
We calculate the effective number of ebits $E_F$ (entropy of formation\cite{Flurin2012, Giedke2003Sep}) at the detectors input, 
which corresponds to a shared number of EPR singlets needed to reconstruct a covariance matrix\cite{Bennett1996} by using the following equation\cite{Flurin2012}: 
\begin{eqnarray}
  E_F &=& c_+ \log_2(c_+) - c_- \log_2(c_-),
  \label{eqn:enform}
\end{eqnarray}
where $c_{\pm} = (\delta^{-1/2} \pm \delta^{1/2})^2/4$ and $\delta=2^{-\mathcal{N}}$.
Now we need to take the bandwidth and logarithmic negativity into account to obtain the potentially available ebit/s of the device.

The photon spectral density is expected to follow a parabolic function\cite{Johansson2010} as a function of frequency:
\begin{eqnarray}
  n(f) &=& n_p \frac{f(f_p-f)}{(f_p/2)^2} \label{eqn:nf},
\end{eqnarray}
where $f$ is the frequency, $n_p = 0.01$ is the peak photon rate and ($f_p$/2) is the pump frequency.
In the measurement we observed a peak logarithmic negativity of $\mathcal{N} = 0.03$ which yields an $E_F = 0.0016$. The effective measurement bandwidth available to us, was limited by surrounding components such as circulators and the Hemt amplifier to 4-8~GHz.
Taking a parabolic spectrum into account and the usable bandwidth of 4~GHz, $\sim 5.2~$~Mebit/s were available to us in this setup.

\begin{figure}[ht]
\centering
\includegraphics{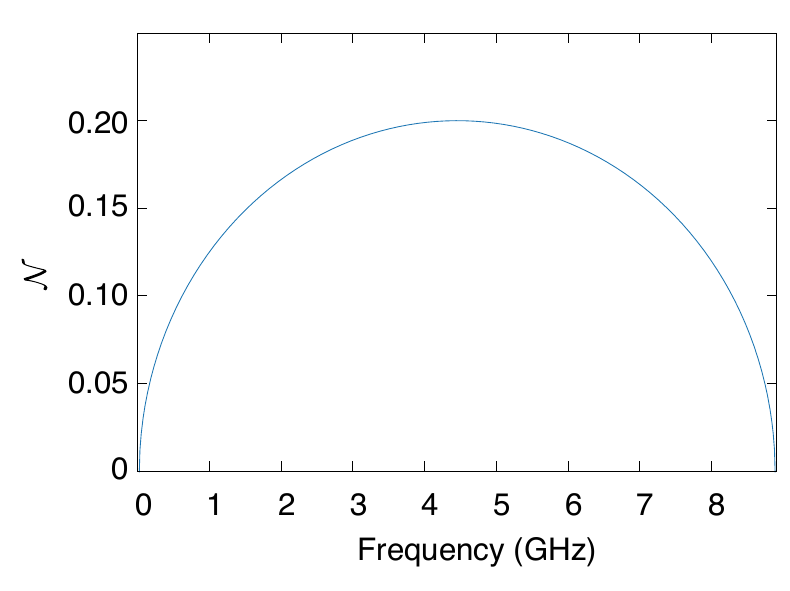}
\caption{Expected logarithmic negativity as a function of frequency $f$.
}
\label{fig:parabola_logN}
\end{figure}

For low photon numbers the theoretical log-negativity is $\mathcal{N} \approx 2 \sqrt{n}$ resulting in:
\begin{eqnarray}
  \mathcal{N}(f) &\approx& 2 \sqrt{\frac{f(f_p-f)}{(f_p/2)^2}},
\end{eqnarray}
which is valid for frequencies between 0 and the pump frequency $f_p$ (Fig. \ref{fig:parabola_logN}).
The integral of this together with the eq. (\ref{eqn:enform}) from 0 to the pump frequency yields $\sim261$~Mebit/s. 
In practice a smaller number is measured due to losses, non-linearity and a limited measurement sensitivity.
In our case, we estimate the losses between HEMT amplifier and sample to be -2.1~dB. Taking this into account, we obtain a log-negativity at the sample of $\mathcal{N} = 0.1$, which corresponds to $\sim 90$~Mebit/s.\\

\begin{table}[ht]\footnotesize
\begin{center}
\begin{tabular}{c | c | c}
\multicolumn{1}{p{2cm}|}{\bfseries Reference} &
\multicolumn{1}{p{3cm}|}{~\bfseries Entanglement rate measured including noise and losses} & \multicolumn{1}{p{3cm}}{~\bfseries Entanglement rate at the sample}\\ 
 \hline
 \cite{Eichler2011} & - & 5.14 Mebit/s \\ 
 \cite{Flurin2012} & - & 6 Mebit/s \\  
 \cite{Menzel2012Dec} & - & 5.7 Mebit/s \\  
 \cite{Ku2015Apr} & 0.07 Mebit/s & 2.7 Mebit/s \\  
 \cite{Fedorov2018Apr} & - & 4.3 Mebit/s \\  
 This work & 5.2 Mebit/s & 90 Mebit/s
\end{tabular}
\end{center}
\caption{Entanglement rate comparison. Here we compare the entanglement rate to other sources for entanglement. The references correspond to the ones in the supplementary.} 
\label{ebit_table}
\end{table}

The estimated entanglement rates presented in table~\ref{ebit_table} are deduced from the two mode squeezing magnitude. This is done by taking losses and noise (including thermal photons) into account for the measured results.
Missing numbers where estimated or obtained from the authors of the papers wherever possible. The calculation including obtained numbers can be found here: \url{https://github.com/benschneider/Dynamical-Casimir-Effect-Sim/blob/master/ebit_comparison.ipynb}.
\pagebreak

\end{document}